\documentclass[%
prd,%
preprint%
,secnumarabic%
,amssymb, amsmath,nobibnotes, aps]{revtex4}
\usepackage{epsfig}%
\usepackage{graphicx}%
\usepackage{amsmath}
\usepackage{amsfonts}
\usepackage{amssymb}
\usepackage{graphicx}

\expandafter\ifx\csname package@font\endcsname\relax\else
 \expandafter\expandafter
 \expandafter\usepackage
 \expandafter\expandafter
 \expandafter{\csname package@font\endcsname}%
\fi

\begin{document}

\title{Relic density and PAMELA events in a heavy wino dark matter model with Sommerfeld
effect}%

\author{Subhendra Mohanty $^a$, Soumya Rao $^a$ and D.P.Roy $^b$ }%
\affiliation{$^a$Physical Research Laboratory, Ahmedabad 380009,
India.
\\$^b$Homi Bhabha Centre for Science Education, Tata Institute of Fundamental
Research,Mumbai-400088, India.
\\
}
\def\be{\begin{equation}}
\def\ee{\end{equation}}
\def\al{\alpha}
\def\bea{\begin{eqnarray}}
\def\eea{\end{eqnarray}}

\begin{abstract}
In a wino LSP scenario
 the
annihilation cross section of winos gravitationally bound in galaxies can be
boosted by a Sommerfeld enhancement factor  which arises due to the ladder of
exchanged $W$ bosons between the initial states. The boost factor obtained can be in
the range $S \simeq 10^4$  if the mass is close to the resonance value of $ M \simeq 4$
TeV.  In this paper we show that if one takes into account the Sommerfeld
enhancement  in the relic abundance calculation then the correct relic density
is obtained for 4 TeV wino mass due to the enhanced annihilation after their
kinetic decoupling. At the same time the Sommerfeld enhancement in the   $\chi
\chi \rightarrow W^+ W^-$ annihilation channel is sufficient to explain the
positron flux seen in PAMELA data without significantly exceeding the observed antiproton
signal. We also show that $(e^- + e^+)$ and gamma ray signals are broadly compatible with the Fermi-LAT observations. In conclusion we show  that a 4 TeV wino DM can explain the positron and
antiproton fluxes observed by PAMELA and at the same time give a thermal relic
abundance of CDM consistent with WMAP observations.

\end{abstract}

\maketitle
\section{Introduction}

There is a great deal of activity in recent years in trying to interpret
the excess of hard positron events reported by the PAMELA experiment
\cite{Pamela-positron}
 as a possible signal for dark matter \cite{Cirelli:2008pk}. In the framework of
the minimal supersymmetric standard models (MSSM), one expects the wino to be
the lightest supersymmetric particle, and hence a dark matter candidate, in the
anomaly mediated supersymmetry breaking model \cite{ Randall:1998uk,
Giudice:1998xp, Gherghetta:1999sw,Wells:2004di} as well as some string inspired
models \cite{Brignole:1993dj,Casas:1996wj} . More over, such a wino dark matter
is known to give the right cosmological relic density for a relatively heavy
wino
mass of ~ 2 TeV \cite{Chattopadhyay:2006xb}. It is also widely recognized that
such a heavy wino dark matter can naturally account for a hard positron signal
from their pair-annihilation into the $W^+W^-$- channel, followed by the
leptonic decay of the W bosons \cite{Hisano:2005ec, Barger-Pamela, Donato:2008jk,
Cholis:2010xb}. It is equally well known, however, that the model would need a
very large boost factor to match the observed magnitude of
the positron signal.

One of the main theoretical mechanisms invoked to explain the above mentioned boost factor
is the so called Sommerfeld effect \cite{Hisano:2005ec, Hisano:2003ec,Profumo,
MarchRussell, ArkaniHamed,Yuan}. It is recognized, however, that in order to get a large
enhancement of DM annihilation cross-section (boost factor), the DM mass has to lie very
close to one of the Sommerfeld resonances \cite{Lattanzi:2008qa,Iengo1,Iengo2}, in which
case the Sommerfeld enhancement will also have a profound effect on the DM relic density.
In this work, we shall assume the wino DM mass to lie very close to the first Sommerfeld
resonance of about 4 TeV, and see if the Sommerfeld effect can simultaneously account for
the right cosmological relic density as well as the large boost factor required to explain
the size of the hard positron signal. Recently a quantitative analysis of the Sommerfeld
effect in the DM relic density calculation has been made by Feng et al \cite{Feng1,Feng2}
in the context of a new physics scenario, where the effect arises from multiple exchanges
of a light gauge boson in a hidden sector. We shall follow their procedure closely for
incorporating the Sommerfeld effect in the relic density calculation for a wino DM of the
MSSM, where it arises from the multiple exchanges of the standard W boson. We shall see
below that for a wino mass of about 4 TeV, the Sommerfeld effect can indeed account for
the right WMAP compatible relic density \cite{Komatsu} as well as a large boost factor,
required to explain the PAMELA positron excess. We shall also see that increasing the wino
mass to 4 TeV helps to make  the predicted antiproton signal compatible with the PAMELA
measurements \cite{Pamela-pbar1,Pamela-pbar2} of $\bar p/p$ ratio  by shifting the
antiproton peak to higher energies. The flux ratio $e^+/(e^+ + e^-)$ measured by PAMELA
\cite{Pamela-positron} gets a contribution from the secondary positrons generated by
cosmic ray electrons and protons. The  measurement of the $(e^+ + e^-)$ flux by Fermi-LAT
\cite{FermiLat-electron1,FermiLat-electron2} along with the measurements of heavier nuclei
flux spectrum in the $(10-600)$ GeV range by HEAO-3 \cite{Engelmann}, ATIC-2 \cite{Panov}
and CREAM \cite{Ahn} fix the parameters of primary cosmic ray background and the diffusion
parameters in the galaxy. This leaves us essentially with the boost factor as the free
parameter to adjust in comparing the theory with the Pamela positron data. Having fixed
the boost factor, which in our case turns out to be $S=10^4$, we show that the
$\gamma$-rays produced by the DM annihilation do not exceed the diffuse $\gamma$-ray
observations by Fermi-LAT\cite{FermiLat-gamma}.

The work is organized as follows. In section 2, we summarize the wino DM model
along with the procedure for incorporating the Sommerfeld effect in the DM relic
density calculation. Then in section 3, we present the results for the  DM relic
density after incorporating the Sommerfeld effect along with the corresponding
boost factor as functions of the DM mass.  In section 4 we compare the model
predictions for the hard positron and antiproton events with the experimental
data. We conclude with a summary of our results in section 5.

\section{Sommerfeld enhancement in wino DM model}

In MSSM with universal gaugino mass $M_{1/2}$ at the GUT scale, the one-loop
renormalized gaugino masses at weak scale  are in the ratio,
\be
M_1:M_2:M_3 :: \alpha_1 :\alpha_2 : \alpha_3 \simeq 1: 2: 7
\ee
and the wino is not the LSP.

The wino LSP scenario is realized in the anomaly mediated supersymmetry breaking
(AMSB) model \cite{Randall:1998uk,Giudice:1998xp,Gherghetta:1999sw,Wells:2004di}
, where the gaugino and scalar masses arise from supergravity breaking in the
hidden sector via super-Weyl anomaly contributions. The gaugino masses are
proportional to the gravitino mass  $m_{3/2}$
\bea
M_1=\frac{33}{5} \frac{\alpha_1}{4 \pi} m_{3/2}, \,\,\, M_2=\frac{\alpha_2}{4
\pi}m_{3/2},\,\,\,M_3=-3 \frac{\alpha_3}{4 \pi} m_{3/2}
\eea
and the renormalised gaugino masses at the weak scale are in the ratio,
\be
M_1:M_2: |M_3| :: \frac{33}{5} \alpha_1 : \alpha_2 : 3 \alpha_3 \simeq  2.8 :
1:7.1
 \ee
In the minimal AMSB model the slepton masses come out negative  at the weak
scale , and this situation is ameliorated by adding a common scalar mass term
$m_0$  at tree level . This model has four free parameters,  $m_{3/2},m_0,
tan\beta$ and $sgn(\mu)$.  Besides the wino turns out to be the LSP in some
string inspired models \cite{Brignole:1993dj,Casas:1996wj}.  Both the wino
annihilation and the Sommerfeld enhancement processes of our interest are
controlled by its isospin gauge coupling $\alpha_2$ to $W$ boson.  So our
results are essentially independent of the underlying SUSY model; and depend
only on the wino LSP mass.

The scattering or annihilation cross section of non-relativistic particles in
the initial state can be substantially changed if there is a long-range force
between the incoming particles which distorts their wave function. The corrected
cross section due to the distortion of incoming states from the plane wave can
be calculated from the wave function of the two-body system in the attractive
potential of the exchanged light  particles. Taking $M$ as the initial particles
mass and $m_\phi$ the exchanged boson mass and $\alpha$ the  coupling,
the L=0 partial wave obeys the Schrodinger equation
\be
\frac{1}{M} \frac{d^2 \psi}{dr^2} + \frac{\alpha}{r} e^{-m_\phi r} \psi= -M v^2
\psi
\label{Schr}
\ee
with the boundary condition $\psi^\prime(r) =i M v \psi(r)$ and $\psi(r)=e^{i M
v r}$ at $r\rightarrow \infty$. The Sommerfeld
enhancement factor is $S=|\psi(\infty)|^2/|\psi(0)|^2$. The Sommerfeld factor
can be calculated by solving the Schrodinger equation (\ref{Schr}) numerically
\cite{ Lattanzi:2008qa, Iengo1,Iengo2} . In this paper we use  an analytical
approximation for the Sommerfeld factor which can be written as \cite{Feng2,
Cassel, Slatyer},
\be
S(v,M,m_\phi)=\frac{\pi  \alpha  \sinh
   \left(\frac{12 M v}{\pi  m_{\phi
   }}\right)}{v \left(\cosh
   \left(\frac{12 M v}{\pi  m_{\phi
   }}\right)-\cos \left(2 \pi
   \sqrt{\frac{6 M \alpha }{\pi ^2
   m_{\phi }}-\frac{36 M^2 v^2}{\pi
   ^4 m_{\phi }^2}}\right)\right)}
\label{S}
\ee

\begin{figure}[h]
\includegraphics[width=12cm]{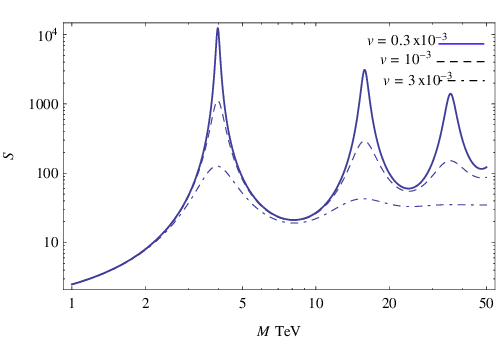}
\caption{ Sommerfield enhancement from $W$ exchange as a function of the  DM
mass for different relative velocities.}
\label{S-v}
\end{figure}

The annihilation cross section of the S-wave initial states gets enhanced at low
velocities by the factor $S$.
For the W ladder processes taking $m_\phi=M_W$  and  $\alpha=\alpha_2=1/30$, the
plots for $S$ as a function of the DM mass $M$ at different velocities is shown
in Fig~\ref{S-v}. We see that the maximum enhancement takes place at $M\sim 4
TeV$ for any given velocity.

The $W$ ladder also gets corrections due to the $\gamma$ and $Z$ exchange between the
chargino intermediate states.  We follow \cite{Lattanzi:2008qa} and ignore the corrections
due to these extra channels which is a adequate approximation owing to the uncertainty in
the knowledge about the velocity distribution of the DM.  We shall see from the relic
density calculation of the next section that the two narrow bands of wino masses
$M=(3.977-3.983)$ TeV and $M=(3.944-3.951)$ TeV are compatible with WMAP measurements at
3$\sigma$.  For the calculation of the boost factor relevant to PAMELA signal we take
$M=3.98$ TeV (at the middle of the higher WMAP compatible band) and assuming the the
galactic rms velocities $v=0.33 \times 10^{-3}$, the Sommerfeld enhancement is $S=10^4$.
The velocity needed for the required boost factor is within the range observed in rotation
curves of galaxies. It is possible that there are cold pockets in the galaxy where the rms
velocities are lower ($v \simeq 10^{-4}$), and in these cold pockets the DM annihilation
can get a larger boost factor ($S \simeq 10^5$) and they can make significant contribution
to the PAMELA signals \cite{Lattanzi:2008qa}.  In that case the boost factor of $10^4$
represents an average DM contribution from such cold clumps and the smooth halo, which is
parameterized here via an effective velocity of $0.33 \times 10^{-3}$.


For the calculation of relic density we use the relation (\ref{S}) for $S(v,M)$ to
calculate the thermal average of the cross-section which is a function of the temperature
$T$ or $x=M/T$, by averaging the cross-section over all velocities at a given temperature,
\be
\langle \sigma v \rangle= \frac{ x^{3/2}}{2 \sqrt{\pi}} {\int_0}^\infty dv
\,(\sigma v)\, S(v,M)\, v^2 \,e^{ -x v^2/4}
\ee
where $v_{rel}=2v$ is the relative velocity between the annihilating particles in the
center of mass frame.  This thermal averaged cross section is used in the calculation of
relic abundance which we describe in the next section.

\section{ Wino relic density with Sommerfeld factor}

The thermal averaged annihilation cross section $\langle \sigma v\rangle(x)$
determines the relic density $n_\chi$ through the Boltzman equation
\be
\frac{dn_\chi}{dt}+3 H n_\chi =- \langle \sigma v\rangle\,
(n_\chi^2-{n_\chi}_{eq}^2)
\ee
where ${n_\chi}_{eq}= g_\chi (M^2/(2 \pi x))^{3/2}\,e^{-x}$ is the the number
density  of $\chi$  produced by the back-reaction
$f f^\prime \rightarrow \chi \chi$  at a given temperature. The Boltzman
equation can be written in terms of dimensionless variables $Y_\chi=n_\chi/s$
(where $s=(2 \pi^2/45) g_{*s}T^3$ is the entropy density) and $x$  as,
\be
\frac{d Y_\chi}{dx} =- \frac{\lambda(x)}{x^2}\,\left( Y_\chi^2(x) -
{Y_\chi}_{eq}^2 (x)\right)
\label{Yeqn}
\ee
where
\be
\lambda(x) \equiv \left(\frac{\pi}{45}\right)^{1/2}\, M\, M_{Pl}
\left(\frac{g_{*s}}{\sqrt{g_*}} \right) \, \langle \sigma v \rangle(x)
\ee
where $g_*$ and $g_{*s}$ are the effective degrees of freedom of the energy density and
entropy density respectively.  The freeze-out temperature is defined as the solution of
\be
Y(x_f)=(1+c) Y_{eq}(x_f)=(1+c) (0.145)\, \frac{g_\chi}{g_{*s}}\, x_f^{3/2} \,e^{-x_f}\quad;
\ee
where the constant  $c\simeq 1$, $g_\chi=2$ for neutralinos; and the freeze-out
temperature turns out to be given by $x_f\simeq 20$.
Below the freeze-out temperature $Y_{eq}(x)$ can be dropped from (\ref{Yeqn})
and we can write the solution of  (\ref{Yeqn}) in the present epoch as
\be
\frac{1}{Y(x_s)}= \frac{1}{Y(x_f)} + \int_{x_f}^{x_s} dx \frac{\lambda(x) }{x^2}
\label{Yxse}
\ee
where $x_s$ is the temperature  where the co-moving density of $\chi$ does not
change noticeably and the integration can be terminated. The DM density fraction
in the present universe is then given by
\be
\Omega=\frac{Y(x_s)\, s_0 \, M}{\rho_{c}}
\ee
where $s_0= 2918{\rm cm^{-3}}$ is the entropy of the present universe and $\rho_{c}=h^2
8.1 \times 10^{-47} {\rm GeV^4} $ is the critical density. The observed value of CDM
density from the seven year WMAP data  is $\Omega_c h^2=0.1123 \pm 0.0105$ (3$\sigma$)
\cite{Komatsu} (where $h$ is the Hubble parameter in units of $100$ km/s/Mpc).

From the calculation of relic density in the wino LSP models
\cite{Chattopadhyay:2006xb} it is known that the WMAP relic density
$\Omega_c h^2=0.1123\pm 0.007 (2\sigma)$ is attained if the wino mass is in the
range $M_2=(2.1- 2.3) {\rm TeV}$. A wino mass of 4 TeV results in  an
overabundance by a factor of 3 larger than the WMAP limit.

In the following we calculate the relic density of WINO LSP by including the
Sommerfeld factor in the annihilation cross section. In  \cite{Hisano:2006nn}
an approximate form of Sommerfeld factor was included  in the annihilation cross
section and it was found that  wino mass which gave the correct relic density
was in the range $M_2=(2.7 -3.0) \rm{TeV}$ .

In this paper we incorporate the Sommerfeld effect following Feng et al
\cite{Feng1,Feng2}, taking into account the fact that the DM particles do not
share the same temperature with radiation bath below a the kinetic decoupling
temperature but in fact cool faster than radiation. The Sommerfeld factor
becomes very effective at temperatures below kinetic decoupling, and that solves
the problem of DM overabundance at M=4 TeV.


For the wino DM, the annihilation channels are $ \chi^0 \chi^0
\overset{\chi^+}{\rightarrow} W^+ W^-$ with a cross section \cite{Hisano:2006nn}
\be
(\sigma v)= \frac{2 \pi\, \alpha_2^2}{M^2}
\ee
and the dominant co-annihilation channel  $ \chi^0 \chi^- \overset{W^-}{\rightarrow} f
f^\prime$ with the cross section for the S-wave initial state
\be
(\sigma v)=\frac{1}{2} \frac{\pi\, \alpha_2^2}{M^2}
\ee
The chargino-neutralino mass difference is $\Delta M \simeq 200 {\rm MeV}$
\cite{Gherghetta:1999sw,Hisano:2004ds}, so the co-annilation channel will not take place
below $\Delta M\simeq (3/2)T$ or $T=133.3 {\rm MeV}$.

{\large\textbf{ Kinetic decoupling:}}
It has long been recognized that after the freezout the dark matter distribution at some
point will not have the same temperature as the radiation bath but cool more
rapidly\cite{Chen, Hofmann,Bringmann}. It has also been pointed out that in cases where
the DM annihilation has strong velocity dependent cross sections as in the case of
Sommerfeld effect the kinetic decoupling of DM will result in rapid annihilation of DM
much below the freeze-out temperature \cite{Dent, Zavala, Feng1,Feng2}. In this section we
calculate the temperature of kinetic decoupling of heavy wino DM and incorporate it in the
wino relic density calculation.

The $\chi_0$ DM will have the same temperature as the radiation bath of
standard model particles till the rate of momentum transfer due to scattering
\be
\Gamma_k= n_r \langle \sigma v \rangle \frac{T}{M}
\ee
is larger than the Hubble expansion rate $H=1.66 \sqrt{g_*}T^2/M_{Pl}$. The kinetic
decoupling temperature $T_{kd}$ is defined as the temperature where
$\Gamma_k(T_{kd})=H(T_{kd})$. For $T> M_W$ the scattering process $ \chi_0 W^+
\overset{\chi^+}\rightarrow \chi_0 W^+$ has a cross section $\langle \sigma v
\rangle= 2 \alpha_2^2/M^2$ and the kinetic decoupling condition gives the
kinetic decoupling temperature to be
\be
T_{kd}^W =2.5 \times \left( \frac{M^3}{\alpha_2^2 M_{Pl}} \right)^{1/2}= 5.4 {\rm
MeV} \left(\frac{M}{4 {\rm TeV}} \right)^{3/2}.
\ee
But at this temperature there are no $W^{\pm}$ in the radiation bath, so the
kinetic decoupling temperature for this interaction channel is $T_{kd}^W=(2/3) M_W=54 {\rm
GeV}$.

Other processes that maintain temperature equality between the DM and the
radiation bath are the elastic scattering with relativistic fermions, $f \chi^0
\rightarrow f \chi^0 $. In the absence of a Higgsino component in the DM there
is no $Z$ exchange diagram, and the  elastic scattering process will come from s
and u-channel sfermion exchanges with the cross section  \cite{Hofmann},
\be
\langle \sigma v \rangle  = 12 \pi  \alpha_2^2 \, I_f^4\,
\frac{E_f^2}{(M_{\tilde f}^2-M^2)^2}
\ee
where $I_f$ is the isospin of $f$ , $M_{\tilde f}$ is the sfermion mass and
$E_f= (3/2)T$ is the fermion energy.
 The kinetic decoupling temperature of the relativistic fermion-wino scattering
process will then be given by
\be
T_{kd}^f= 4.2 \left ( \frac{(M_{\tilde f}^2-M^2)^2 M }{ M_{Pl}} \right)^{1/4}
\ee
If we take $M=4{\rm TeV}$ and $M_{\tilde f} = 10 {\rm TeV}$, then the kinetic
decoupling for this channel takes place at $T_{kd}^f= 5.2{\rm GeV}$ while for
$M_{\tilde f}=5 {\rm TeV},T_{kd}^f= 1.7 {\rm GeV} $.

\begin{figure}[h!]
\includegraphics[width=10cm]{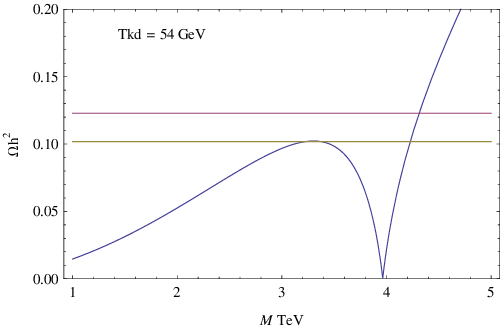}
\includegraphics[width=10cm]{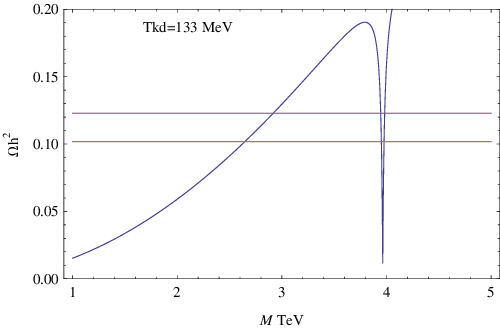}
\caption{Wino relic density with Sommerfeld enhancement with two
different kinetic decoupling temperatures as discussed in the text. $T_{kd}=133$MeV
(bottom panel) is assumed for the wino relic density calculation. The $3\,\sigma$ band of
WMAP relic density \cite{Komatsu} is indicated by the horizontal lines.}
\label{Tkd}
\end{figure}

Finally the quasi-elastic scattering $\chi^0 f \overset{W}{\leftrightarrow} \chi^\pm
f^\prime$ will maintain the kinetic-coupling of $\chi^0$ with radiation down to
a temperature $T=(2/3)\Delta M=133{\rm MeV}$, assuming a typical charged and
neutral wino mass difference $\Delta M=200$ MeV \cite{Gherghetta:1999sw,
Hisano:2004ds}. Below the temperature $133 {\rm MeV}$ there will be no charginos
in the heat bath so the kinetic decoupling of the winos from the heat bath  will
take place at $T_{kd}= 133 {\rm MeV}$. Since this is the lowest of the kinetic
decoupling temperatures from various processes discussed above, we will take
$T_{kd}=133 {\rm MeV}$ for the calculation of wino relic abundance. In
Fig~\ref{Tkd} we show the effect of the kinetic decoupling temperature on the
relic density. Early de-couplings enhance the
annihilation and lower the relic abundance. We take $T_{kd}=133$MeV as the
effective kinetic decoupling temperature for calculating the relic abundance.

\begin{figure}[h!]
\includegraphics[width=0.8\textwidth]{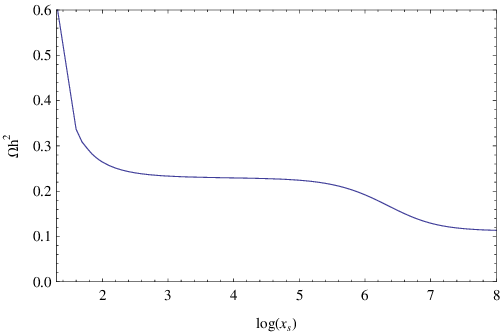}
\caption{Relic density as a function of cutoff temperature $x_s$ for a wino DM mass M=4 TeV. There is a
fall at $x_s=10^5$ due to kinetic decoupling, and the chemical decoupling occurs
at $x_s=4 \times 10^7$. }
\label{Yxs}
\end{figure}

At temperatures below $T_{kd}$ the temperature of the decoupled non-relativistic particles
is related to their momenta as $(3/2)T=M v^2/2$. As the velocities  of NR particles in the
expanding universe goes down with the scale factor as $v \propto 1/a $,   DM temperature
falls with the scale factor as $1/a^2$ while the radiation temperature goes as $1/a$. The
integration in (eqn. \ref{Yxse}) in the interval $x> x_{kd} \equiv M/T_{kd}$ is performed
by replacing $x$ in the integrand by $x^2/x_{kd}$, which corresponds to the temperature of
the DM particles. This is because the thermal equilibrium in this region is maintained by
scattering between DM particles via multiple W boson exchange (Sommerfeld ladder) i.e $\chi^0 \chi^0 \overset{W}\leftrightarrow \chi^+ \chi^-$. Thanks to the Sommerfeld effect this process can continue for $T<T_{kd}$(133 MeV). The
rapid reduction in velocities or temperatures of DM particles below the kinetic decoupling
temperatures results in a large enhancement in the annihilation cross section and
reduction of the relic abundance in the late universe ($T < T_{kd}$).  With these inputs
we carry out the integration in equation (\ref{Yxse}) numerically to obtain $Y(x_s)$. We
plot $Y(x_s)$ in Fig~\ref{Yxs} and find that $Y(x_s)$ remains constant after $x_s > 4
\times 10^7 $ which corresponds to the temperature $\sim 100 keV$. Below this temperature
there is a chemical decoupling of the winos in the sense that their comoving abundance
remains constant. This can be physically understood as follows.  In the range $x_{kd}< x<
x_s$, the DM particles are kept in thermal equilibrium by their mutual scattering $\chi^0 \chi^0 \overset{W} \leftrightarrow\chi^+ \chi^-$. But at
$x>x_s$ the scattering rate for this process falls below the Hubble expansion; so the DM particles no
longer maintain a Maxwell-Boltzmann thermal distribution. Consequently the low velocity
particles with enhanced annihilation rate (\ref{S}) are decoupled from the high velocity
particles. The latter escape annihilation, leading to a constant comoving abundance of DM
at $x>x_s$ and this process is called chemical decoupling.

\begin{figure}[h!]
\includegraphics[width=0.8\textwidth]{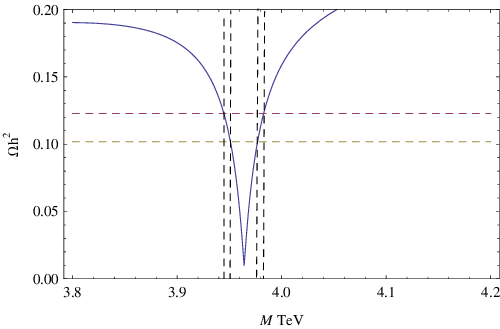}
\caption{Plot shows the relic density as a function of wino mass close
to the Sommerfeld resonance. We see that two  narrow bands of wino masses
$M=(3.977-3.983)$TeV and $M=(3.944-3.951)$TeV are compatible with WMAP
measurements at 3$\sigma$.} 
\label{Omega-M}
\end{figure}

In Fig~\ref{Omega-M} we show the relic density as a function of wino mass close
to the Sommerfeld resonance. We see that two  narrow bands of wino masses
$M=(3.977-3.983)$TeV and $M=(3.944-3.951)$TeV are compatible with WMAP
measurements at 3$\sigma$. Admittedly there is fine-tuning involved in reproducing the WMAP relic densitywith the Sommerfeld resonance. In the absence of a unique measure of fine-tuning, however, we have not given a numerical esimate of this quantity. Instead we have presented this fine-grained plot from which the interested reader can estimate the
fine-tuning measure of his choice.

\begin{figure}[h!]
\includegraphics[width=0.8\textwidth]{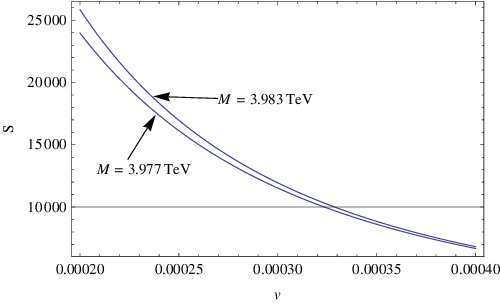}
\caption{Sommerfeld enhancement $S$ for the range of M = (3.977-3.983) TeV (compatible with WMAP relic density) as a function of velocity. A Sommerfeld
enhacement factor of $S=10^4$ is obtained for $v=0.33 \times 10^{-3}$ and M=3.98TeV. }
\label{S-M-v}
\end{figure}

Fig.~\ref{S-M-v} shows Sommerfeld enhancement $S$ for the range of M = (3.977-3.983) TeV
as a function of velocity. A Sommerfeld enhacement factor of $S=10^4$ is obtained for
$v=0.33 \times 10^{-3}$ and M=3.98TeV.

We choose the central values $M=3.98$TeV (which is the center of the higher mass
band)  to compute the positron, electron, antiproton and diffuse $\gamma$-ray flux to compare with
the observations of PAMELA and Fermi-LAT .

\section{Comparison with PAMELA and Fermi-LAT data}
The satellite based PAMELA experiment measures the flux spectrum of $e^+/(e^+ +e^-)$
\cite{Pamela-positron} in the energy range of $(10-100)$ GeV and $\bar p/p$
\cite{Pamela-pbar1, Pamela-pbar2} in the range $(1-180)$GeV. It is found that the
$e^+/(e^+ +e^-)$ ratio exceeds the flux estimated  from astrophysical sources by a large
margin at higher energies.  However the $\bar p$ flux is within the expected range of what
is expected of secondary $\bar p$ produced from the primary CR protons from AGN's and
other sources. Any DM annihilation or decay model must explain the large flux of positrons
and the paucity of $\bar p$'s in the PAMELA signal. In addition there are measurements of
$(e^+ +e^-)$ and $\gamma$-rays in the by Fermi-LAT
\cite{FermiLat-electron1,FermiLat-electron2, FermiLat-gamma}, and the DM  signal must be
consistent with the Fermi-LAT data \cite{Barger,
Palomares,Bernal,Cirelli2,Rizzo,Cholis,Ishiwata}. In this section we check the predictions
of the heavy wino DM model with the PAMELA and Fermi-LAT data.

We use the  MicrOMEGAs \cite{Belanger:2008sj} to calculate rate of  of
positrons and antiprotons produced by the annihilation of $3.98$ TeV wino DM. As
input we take the following SUSY parameters $\mu= 9$ TeV,  $M_2=3.837$ TeV  and
the other gauginos in the ratio $M_1:M_2:M_3=2.8:1:7.1$  while the squark and
slepton masses $M_{\tilde q}=M_{\tilde f}=10$ TeV. We take $tan\beta=10$ and
$\mu=9$TeV.  As mentioned earlier, however, the results are insensitive to all
the SUSY parameters other than the LSP mass. We calculate the annihilation cross section of a 3.98 TeV wino LSP
 and the branching spectrum $(\frac{dN}{dE})$ into electrons, positrons, antiprotons and photons using MicroOMEGAs.
We fit the positron and antiproton spectrum from MicrOMEGAs  with an analytical
functions which is then input in the GALPROP code to calculate the observed fluxes.

\begin{figure}[h!]
 \includegraphics[width=8cm]{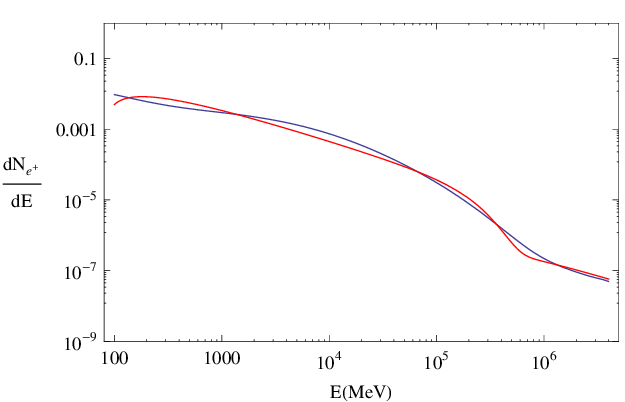}
 \includegraphics[width=8cm]{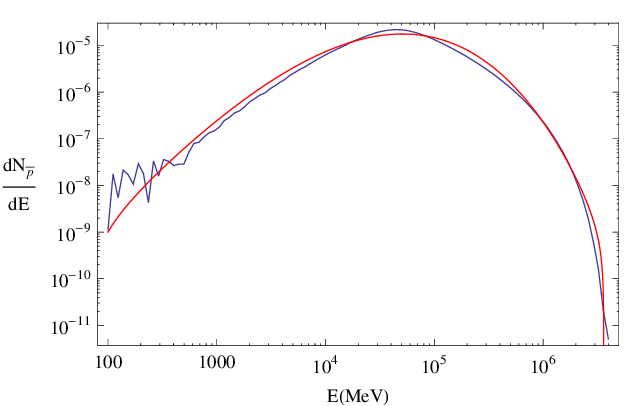}\\
 \centering
 \includegraphics[width=8cm]{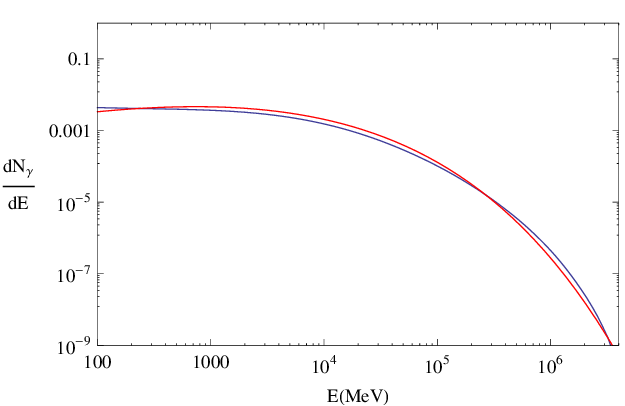}
 \caption{Top left panel shows positron spectrum, top right panel shows
 antiproton spectrum and bottom panel shows gamma ray spectrum from wino
 annihilation.  Light shaded curve shows the fitted functions (\ref{dnpos},
 \ref{dnpbar}) and \ref{dngamma}.}
 \label{Fits}
\end{figure}

  For electrons/positrons we use
\be
\left(\dfrac{dN_e}{dE_e}\right)=E_e^{-1}\left[5\exp\left[-\left(0.018\dfrac{E_e}
{M_\chi}
 \right)^{1.5}\right]-\left(\dfrac{E_e}{ M_\chi }\right)^{ -0.4 }+0.3\right],
 \label{dnpos}
 \ee
 while for protons/antiprotons the we use the analytical fit,
 \be
 \left(\dfrac{dN_p}{dE_p}\right)=E_p^{-1}\left[5.5\left(\dfrac{E_p}{M_\chi}
\right)^4\exp
 \left[ -6.35\left(\dfrac{E_p}{ M_\chi }\right)^{0.25}\right]-3\times 10^{-6}
 \left(\dfrac{E_p}{ M_\chi }\right)\right]
 \label{dnpbar}
 \ee
 and for $\gamma$-ray spectrum we use
 \begin{equation}
  \left(\dfrac{dN_\gamma}{dE_\gamma}\right)=E_\gamma^{-1}\left[2000\left(
  \dfrac{E_\gamma}{M_\chi}\right)^{1.8}\exp\left[-5\left(
  \dfrac{E_\gamma}{M_\chi}\right)^{0.24}\right]\right]
  \label{dngamma}
 \end{equation}

where $E_e$ , $E_p$ and $E_\gamma$ are in units of MeV, while $M_\chi \simeq 4$ TeV. The
integrated number of electrons/positrons, antiprotons and photons per wino pair annihilation are
$N_e= 14.2$ , $N_p=1.59 $ and $N_\gamma=26$ respectively. We show the result of the output from
MicrOMEGAs along with our analytical fits (\ref{dnpos},\ref{dnpbar}) and \ref{dngamma} in
Fig~\ref{Fits}.

The cross section from MicrOMEGAs of the wino annihilation to $W^+ W^-$ is $5.63 \times
10^{-27} cm^3/s$.   We use the Galactic propagation code GALPROP
\cite{Moskalenko:1997gh,Strong:2007} to obtain the positron and antiproton fluxes at the
earth. We input a boosted cross section $(\sigma v)_0S= 5.63 \times 10^{-23} cm^3/s$ in
the GALPROP code along with the analytical fits of $(dN/dE)$ (\ref{dnpos},\ref{dnpbar}).
We choose the isothermal dark matter distribution profile \cite{IsoT} and calculate the
flux of $e^{+},e^{-},p,\bar p$ on earth arising from the wino annihilation in the galaxy.
\begin{table}
 \centering
 \begin{tabular}{|c|c|c|c|c|c|c|c|c|c|}
  \hline\hline
  $L$ & $D_{0xx}$ & $\delta$ & $V_a$ & $\frac{\partial  V_c}{\partial z}$ & $\gamma_n$ & $\gamma_e$ & $N_p$ at 100 GeV & $N_e$ at 34.5 GeV&S\\
  (kpc) & $(10^{28}$ cm$^2)$ & & (km/s) & (km/s/kpc) & & & MeV$^{-1}$cm$^{-2}$s$^{-1}$sr$^{-1}$
  & MeV$^{-1}$cm$^{-2}$s$^{-1}$sr$^{-1}$&
   \\\hline
  2.0 & 2.83 & 0.34 & 33.67 & 0.5 & 2.36 & 2.5 & $3.5\times 10^{-9}$
  & $0.4\times 10^{-9}$&10000\\
  \hline\hline
 \end{tabular}
 \caption{Diffusion parameters and boost factor used as input in GALPROP. }
 \label{difpar}
\end{table}

\begin{figure}[h!]
\includegraphics[width=12cm]{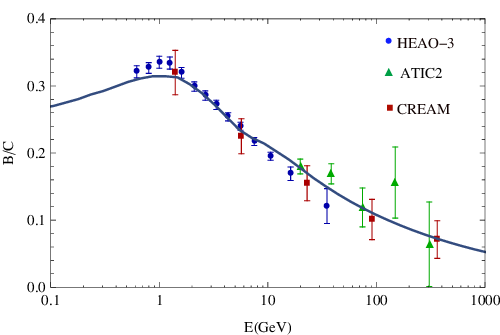}
\caption{Boron/Carbon ratio for the assumed galactic diffusion parameters shown in Table 1. compared with data from HEA0-3 \cite{Engelmann}, ATIC-2 \cite{Panov} and CREAM \cite{Ahn} experiments.}
\label{bcrat}
\end{figure}

These inputs from MicroOmegas goes into the the source term
\be
q(\vec{r},p)=\langle \sigma v\rangle_0 \,S\, \frac{\rho}{M^2}\, \left(\frac{dN}{dE} \right)
\ee
of the propagation equation for the cosmic ray density (at galactic radius $r$ and with
momentum $p$) \cite{Strong:2007},
\begin{equation}
 \frac{\partial\psi(\vec{r},p)}{\partial t}=q+\vec{\nabla}\cdot
 \left(D_{xx}\vec\nabla\psi-\vec{V_c}\psi\right)+\frac{\partial}{\partial p}p^2
 D_{pp}\frac{\partial}{\partial p}\frac{1}{p^2}\psi-\frac{\partial}{\partial p}
 \left[\dot p\psi-\frac{p}{3}\left(\vec\nabla\cdot\vec{V_c}\right)\psi\right]-
 \frac{1}{\tau_f}\psi-\frac{1}{\tau_r}\psi
\end{equation}
where  $D_{xx}$ is the diffusion coefficient which describes the scattering of CR by the
random galactic magnetic fields, $V_c$ is the convection velocity of the bulk CR in the
galaxy (and this term represents the scattering of the CR by the background CR "wind"),
$D_{pp}$ is the diffusion coefficient in momentum space (which represents acceleration in
turbulent B fields), $\dot p$ is the energy loss due to radiative decay and the final two
terms represent the possible fragmentation or radioactive decay of the CR nuclei. This
equation is solved over a diffusion zone represented by a cylinder whose origin and axis
coincide with our galactic disk.  The half height of this cylinder is typically $L\sim
1-10$ kpc \cite{Rizzo}.  The energy dependent diffusion coefficient is parameterized as
$D_{xx}(\vec r,E)=D_{0xx}E^\delta$.  The primary spectrum for all nuclei is parameterized
as\cite{Rizzo} \begin{equation} \psi = \frac{N}{2}\frac{L}{D_{0xx}}E^{-\gamma_n-\delta}
\end{equation} and a similar expression for electrons with $\gamma_n$ replaced by
$\gamma_e$. For a given set of input parameters characterizing  the primary spectra for
any CR species, the GALPROP code propagates the CR flux and gives the flux observed at the
earth.   We choose the set of parameters $D_{0xx},L, V_c, \delta, V_a ({\rm Alfven \,
velocity}),\frac{\partial  V_c}{\partial z},\gamma_n $ (tabulated in Table.\ref{difpar})
which gives the Boron/Carbon ration in CR consistent with measurements by HEAO-3
\cite{Engelmann}, ATIC-2 \cite{Panov} and CREAM \cite{Ahn} experiments as shown in
Fig\ref{bcrat}. This is not the only parameter set which gives the correct B/C spectrum,
but in our scan of parameters we find that this set gives a smallest contribution to the
$\bar p/p$ flux ratio.

\begin{figure}[h!]
\includegraphics[width=12cm]{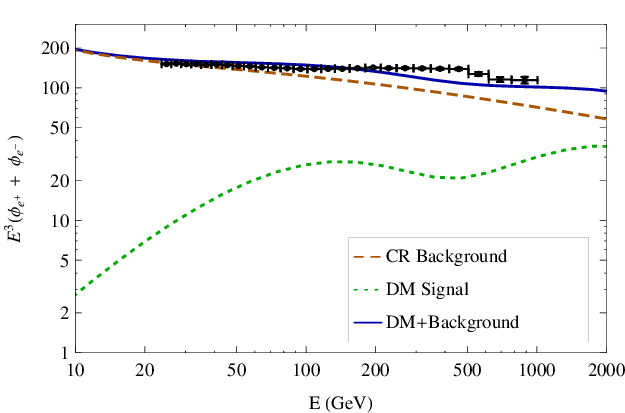}
\caption{The $(e^- + e^+) $ flux  for the 3.98 TeV wino DM compared with FERMI-LAT data
\cite{FermiLat-electron1, FermiLat-electron2}. Dashed denotes the CR background and dotted line is the DM annihilation signal.}
\label{fermi-electron}
\end{figure}

The primary-electrons spectral index $\delta_e=2.5$ is chosen to fit the Fermi $(e^{+} +
e^{-})$ data. The primary proton flux $N_p$ and electron flux $N_e$ are also fixed by
fitting with the $(e^+ + e^-)$ and $B/C$ observations. This set of parameters fixes the
background $\bar p/p$, $e^{-}+e^{+}$ and the $\gamma$-ray photon. In
Fig.\ref{fermi-electron} we show the $(e^+ + e^-)$ signal from the $3.98$ TeV wino DM
model with observations from Fermi-LAT \cite{FermiLat-electron1, FermiLat-electron2}.

With the parameters for the background thus fixed we vary the boost factor
trying to get a fit for the $(e^+/(e^+ + e^-)$ data while at the same time being
consistent with the $\bar p/p$ data. We find that the boost factor $S=10^4$
gives
a good fit to the PAMELA positron data with a $\chi^2/d.o.f= 1.03$ while being
consistent with the PAMELA antiproton data (with $\chi^2/d.o.f=1.68$).  The large DM mass
helps to ameliorate the discrepancy with the PAMELA antiproton data by pushing up the
predicted peak to still higher energies.

\begin{figure}[h!]
\includegraphics[width=12cm]{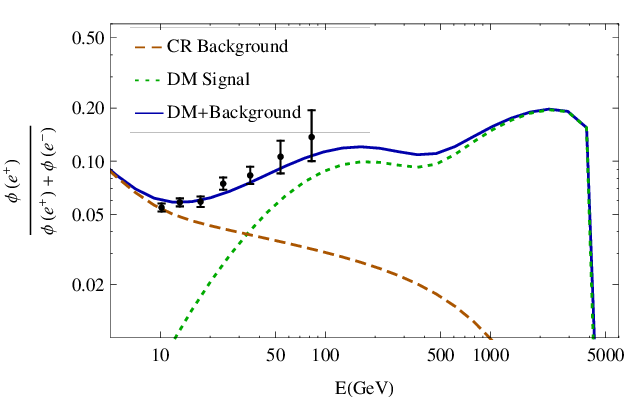}
\caption{Positron flux ratio for the 3.98 TeV wino DM compared with  Pamela data
\cite{Pamela-positron}. Dashed line shows background from cosmic ray secondary
positrons.}
\label{pos}
\end{figure}

\begin{figure}[h!]
\includegraphics[width=12cm]{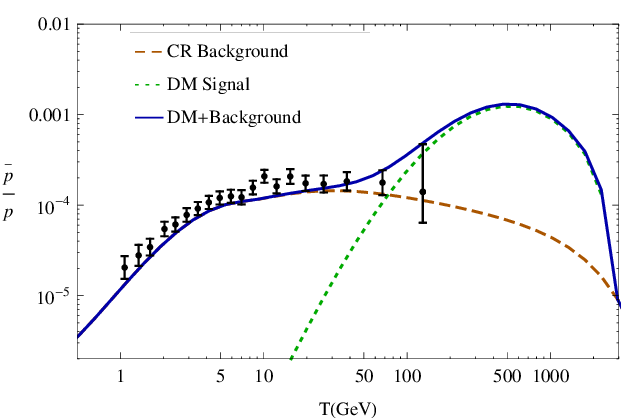}
\caption{Antiproton/Proton flux ratio  for the 3.98 TeV wino DM compared with Pamela data
\cite{Pamela-pbar1,Pamela-pbar2}. Dashed line shows background from cosmic ray secondary
antiprotons while  dotted line shows DM signal.}
\label{pbar}
\end{figure}

The total flux of positrons and antiprotons from  the background as above plus
the DM signal taken from GALPROP is shown in Fig~\ref{pos} and Fig~\ref{pbar}
respectively.


\begin{figure}[h!]
	\centering
\includegraphics[width=12cm]{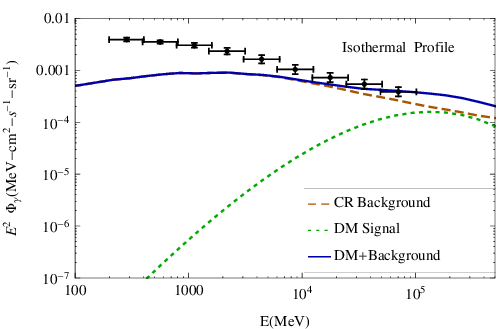}
\includegraphics[width=12cm]{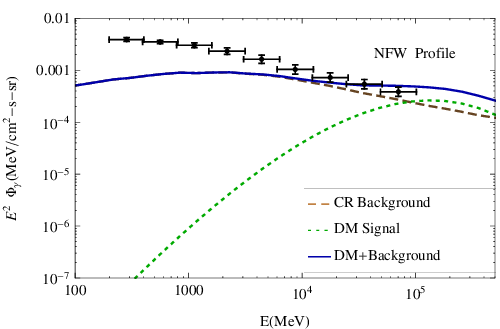}
\caption{Diffuse $\gamma$-ray flux for the 3.98 TeV wino DM compared with
FERMI-LAT integrated $\gamma$-ray flux at galactic latitudes $b>20^\circ$
\cite{FermiLat-gamma}. Dashed line shows background. The prompt photon signal is smaller
in the isothermal profile (top panel) compared to the NFW profile (bottom panel).  At
lower energies the primary contribution to the $\gamma$-ray flux is  from astrophysical
sources like  GRB's, gamma ray pulsars and AGN's which is beyond the scope of the present
work. }
\label{fermi-gamma}
\end{figure}

The $\gamma$-ray spectrum observed at earth from DM annihilation photons is given by
\be
\Phi_\gamma= \frac{1}{4 \pi} \,\langle \sigma v \rangle_0\, S\, \frac{dN_\gamma}{dE_\gamma} \int_\Delta \Omega d\Omega \int_{los} \rho^2(l) dl
\ee
where $\rho(l)$ is the DM density along the line of sight (los). In addition to the prompt
photons from DM, photons are generated also by synchrotron radiation, inverse-Compton
scattering from the CMB and infrared starlight from the primary electrons. The prompt
photons are sensitive to the large scale DM distribution function $\rho(l)$ along the line
of sight. We find that the Isothermal distribution \cite{IsoT} gives a smaller gamma-ray
flux compared to other DM profiles like NFW \cite{Navarro:1995iw}, Moore
\cite{Diemand:2005wv} or Einasto \cite{Navarro:2008kc}. We plot the diffuse $\gamma$-ray
flux for the $M=3.98$TeV wino model  for both the NFW ,
\be
\rho_{NFW}=\rho_0 \frac{r_s}{r} \left(1+\frac{r}{r_s}\right)^{-2}, \quad  \rho_0=0.26 {\rm
GeV/cm^3}, r_s=20{\rm kpc},
\ee
as well as the isothermal profiles
\be
\rho_{ISO}=\rho_{\odot}  \frac{1+(\frac{r_{\odot}}{r_s})^2}{1+(\frac{r}{r_s})^2}, \quad
\rho_{\odot}=0.3 {\rm GeV/cm^3},r_{\odot}=8.5{\rm kpc}, r_s=5 {\rm kpc},
\ee
The local DM density for both profiles is $ \rho_{\odot}=0.3 {\rm GeV/cm^3}$ and $r_s$ is
the size of the DM halo. In Fig.\ref{fermi-gamma} we show the plots for the integrated
$\gamma$-ray signal at large galactic latitudes $b>20^\circ$ observed  by Fermi-LAT
\cite{FermiLat-gamma}.The  $\gamma$-ray signal from $3.98$ TeV wino annihilation is
consistent with measurements from Fermi-LAT. At lower energies the primary contribution to
the $\gamma$-ray flux is  from astrophysical sources like  gamma ray blazars and
intergalactic shock waves from structure formation \cite{FermiLat-gamma2}, which is beyond
the scope of the present work.


\begin{figure}[h!]
\centering
\includegraphics[width=0.75\textwidth]{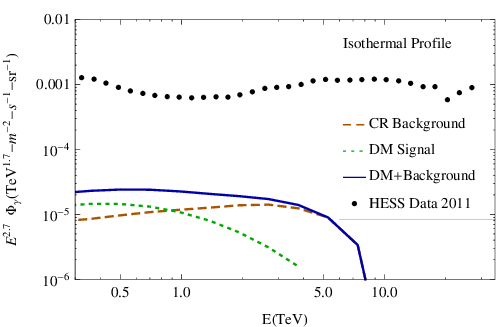}
\includegraphics[width=0.75\textwidth]{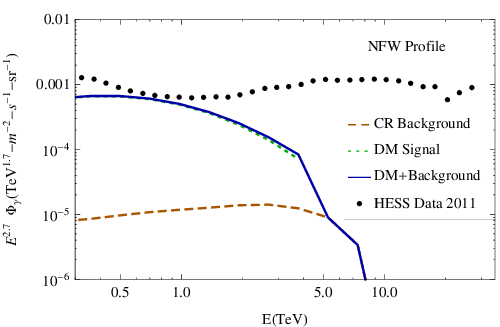}

\caption{Diffuse $\gamma$-ray flux for the 3.98 TeV wino DM compared with HESS
\cite{Abramowski:2011hc} integrated $\gamma$-ray flux from a circular region of $1^\circ$
centred around the galactic centre(GC).  The DM signal from the GC is much smaller in the
isothermal profile(top panel) compared to the NFW profile(bottom panel).}
\label{hess}
\end{figure}

Recent results from HESS reported diffuse $\gamma$-ray flux measurements from the galactic
centre \cite{Abramowski:2011hc}.  We show the effect of the Isothermal and NFW profiles on
the results from HESS in Fig.~\ref{hess}.  For both these profiles the DM signal is in the
acceptable range although for NFW it is much larger than that for the Isothermal profile.
This is due to the fact that the NFW profile is more sharply peaked at the galactic centre compared to
the isothermal profile.

In this paper we have taken into account the non-perturbative effect of the W exchange in
the initial state which results in the Sommerfeld enhancement in the cross section. It has
been seen that the electroweak corrections in the final states and the bremsstrahlung of W
and Z from the external legs in TeV scale DM annihilation can increase the cross section
by a factor $(1+ \alpha_2 \ln^2(M^2/M_W^2)$ \cite{Ciafaloni,Bell}. This increase in the
effective cross section by  $\sim 30\%$ can be broadly taken into account in our analysis
by reducing the necessary Sommerfeld enhancement by a corresponding factor of $1.3$.
There are interesting signals of Sommerfeld enhancement of the DM annihilation cross
section in the CMB anisotropy \cite{cmb1,cmb2}, and they can be pursued  as  potential
signals of heavy wino DM which may be observed in the PLANCK \cite{planck} CMB anisotropy
measurements.

\section{Conclusions}
We have investigated the possibility of reconciling the predictions of a heavy wino dark
matter model with the WMAP data on relic density as well as the hard positron signal
reported by the PAMELA experiment via the Sommerfeld enhancement factor. We find that they
can be simultaneously explained if one assumes the wino mass to be very close the
Sommerfeld resonance mass of about 4 TeV. In that case the large Sommerfeld enhancement of
dark matter annihilation cross-section below the freeze-out temperature reduces the
present dark matter relic density to the range of the WMAP measurement. Moreover the
Sommerfeld enhancement can boost the present dark matter annihilation cross-section by a
large factor of about $10^4$, as required to explain the size of the PAMELA positron
signal. At the same time the large DM mass helps to ameliorate the discrepancy with the
PAMELA antiproton data by pushing up the predicted peak to still higher energies.  We also
find that the model predictions for $(e^+ + e^-)$ and gamma-ray signals are broadly
compatible with the Femi-LAT observations.

\section{Acknowledgements}
We thank Manuel Drees for many helpful comments and discussions.  DPR
acknowledges partial financial support from INSA under the senior scientist scheme. This
work was started as a working group project during WHEPP XI at PRL, Ahmedabad and advanced further
during the NIUS Summer camp at HBCSE(TIFR),Mumbai.

\end{document}